\UseRawInputEncoding
\documentclass[%
 reprint,
 amsmath,amssymb,
 aps,
floatfix,
]{revtex4-2}

\usepackage{graphicx}
\usepackage{dcolumn}
\usepackage{bm}
\usepackage{braket}
\usepackage{placeins}
\usepackage{listings}
\usepackage{xcolor}
\usepackage{float}
\usepackage{tikz}
\usepackage{url}
\usepackage{soul}
\usetikzlibrary{arrows.meta, positioning, calc, fit, backgrounds, decorations.pathreplacing}


\lstdefinestyle{pythonstyle}{
    language=Python,
    basicstyle=\ttfamily\footnotesize,
    keywordstyle=\color{blue},
    commentstyle=\color{gray},
    stringstyle=\color{red},
    showstringspaces=false,
    breaklines=true,
    frame=single,
    tabsize=4,
    captionpos=b
}
\usepackage[colorlinks=true,urlcolor=blue,citecolor=blue,linkcolor=blue]{hyperref}

\begin{document}

\preprint{APS/123-QED}

\title{Physics-Based Flow Matching for Full-Field
Prediction of Silicon Photonic Devices}

\author{Joseph Quaratiello\textsuperscript{1} and Anthony Rizzo\textsuperscript{1}}

\address{\textsuperscript{1}Thayer School of Engineering, Dartmouth College, Hanover, NH 03755}

\date{\today}

\begin{abstract}
  Designing photonic integrated circuits requires accurate electromagnetic field simulations, which remain computationally expensive even for simple device geometries. We present PIC-Flow, a generative neural surrogate that predicts electromagnetic field distributions for photonic devices given their geometry and operating wavelength as an alternative to costly finite-difference time-domain (FDTD) simulations. Our approach combines three key ideas: (i) conditional flow matching as the generative framework, learning a velocity field that transports Gaussian noise to physically valid field solutions; (ii) a real-valued U-Net operating on split real and imaginary field channels; and (iii) physics-constrained training through a Helmholtz residual loss enforcing $\nabla^2 E_z + k_0^2 \varepsilon E_z = 0$. We introduce an interface-aware masking scheme for the Helmholtz residual that excludes dielectric boundary pixels where finite-difference stencil errors dominate, yielding a physically meaningful compliance metric. The data set consists of 22,500 ground-truth FDTD simulations split evenly between multimode interferometers, Y-branches, and directional couplers at $\lambda=1.55\,\mu$m in an 80/10/10 split between training, validation, and test sets. We evaluate ablations on the network against the held out test devices and also show that the model generalizes to held out device classes such as S-bends, tapers, and cascaded Y-branches. Rather than a drop-in replacement for FDTD, this work establishes a foundation that, with broader data coverage, more compute, and further training optimization, could scale toward broadband, device-agnostic field prediction with dramatically improved runtime for rapid design-space exploration of complex photonic devices and circuits.
\end{abstract}

\maketitle


\section{\label{sec:level1}Introduction}

Silicon photonic devices used in photonic integrated circuits (PICs) enable compact optical systems for communications~\cite{sun2015single, rizzo2023massively, daudlin2025three}, sensing~\cite{rogers2021universal, vos2007silicon, yu2018silicon}, and computing~\cite{ahmed2025universal, hua2025integrated, bandyopadhyay2024single}, but practical design still depends on full-wave electromagnetic simulation and iterative optimization. Finite-difference time-domain (FDTD) solvers remain the main tool for accurate prediction across device geometries and wavelengths~\cite{TafloveHagness2005FDTD, Teixeira2023FDTDMethods}, yet repeated FDTD simulations become a bottleneck for parameter sweeps and gradient-based inverse-design loops. This has motivated machine-learning surrogates for forward prediction and learning-assisted inverse design. A key remaining challenge is accurate full-field prediction of complex electromagnetic responses: phase must be handled consistently, solutions should satisfy the wave equation, and models must generalize across device families rather than overfitting to one geometry class. Predicted fields should also preserve device-level behavior such as port-to-port transmission, reflection, and splitting ratios, the quantities ultimately used in photonic circuit design.

We address this gap with a physics-based conditional flow-matching model for fast full-field prediction of silicon photonic devices. Building on flow matching for generative modeling~\cite{Lipman2023FlowMatching} and physics-based flow matching for PDE-constrained generation~\cite{Baldan2025PBFM}, we train a real-valued U-Net to generate the real and imaginary components of the electromagnetic field. Physics is enforced through a Helmholtz residual loss with an interface-aware mask that excludes dielectric boundary pixels where finite-difference stencil errors dominate, yielding a meaningful compliance metric. We study three matched training objectives: flow matching only, flow matching plus phase loss, and flow matching plus both phase and Helmholtz residual losses. This isolates how data fidelity, phase alignment, and PDE compliance affect full-field accuracy. We evaluate the approach on multimode interferometers, Y-branches, and directional couplers, producing full-field predictions in a single forward pass significantly faster than standard FDTD simulations.

This work makes four main contributions: (i) a flow-matching neural surrogate that predicts the complex $E_z$ field of parameterized silicon photonic devices conditioned on permittivity, source mask, and wavelength; (ii) a masked Helmholtz residual loss that excludes PML, source, and dielectric-interface pixels, defining a physically meaningful per-sample compliance metric ($\rho_R$, Eq.~\eqref{eq:compliance}); (iii) an ablation showing that combining flow matching with phase and Helmholtz residual losses improves wave-equation compliance on held-out test samples without sacrificing reconstruction quality, together with a wall-clock benchmark against multi-threaded CPU FDTD on the same compute node; and (iv) evidence that the trained model generalizes to held-out device classes (S-bends, long tapers, cascaded $1\times 3$ Y-branches) that were intentionally omitted from the training data set. This work establishes a proof-of-principle foundation that can dramatically accelerate design cycles for PIC development relative to current standards.

\section{\label{sec:related}Related Work}

Our work connects three lines of research in the domain of computational electromagnetics and machine learning: learning-based surrogates for electromagnetic simulation, physics-informed or physics-constrained training for PDE-governed systems, and generative models for scientific computing. In this section, we provide a brief survey of the current state-of-the-art and discuss the novel contributions of PIC-Flow relative to previous demonstrations in each domain. We additionally briefly discuss inverse design and its relationship to PIC-Flow.

\paragraph{Neural surrogates for photonic simulation.}
Deep learning has been used to approximate forward electromagnetic responses and accelerate design-space exploration across many photonic structures. Jiang \emph{et al.}\ survey deep neural networks for evaluation and design, covering supervised surrogates, inverse models, and hybrid workflows~\cite{Jiang2021DeepNNPhotonicDevices}; Ma \emph{et al.}\ provide a broader review of deep learning for photonic structure design~\cite{Ma2021DLPhotonicsReview}. In integrated photonics, Tahersima \emph{et al.}\ demonstrate data-driven inverse design of compact power splitters~\cite{Tahersima2019DNNSplitters}, and Tang \emph{et al.}\ introduce a generative model for inverse design of nanophotonic components~\cite{Tang2020GenerativeLPR}. Chen \emph{et al.}\ train a physics-augmented U-Net~\cite{Ronneberger2015UNet} (WaveY-Net) to predict near-fields of freeform nanophotonic devices, using Maxwell's equations as part of the training loss~\cite{Chen2022WaveYNet}, and Lim and Psaltis use Maxwell's equations directly as the sole training objective for a physics-driven DNN~\cite{Lim2022MaxwellNet}. These works show the value of learned surrogates, but they also highlight the need for models that predict \emph{full-field} complex solutions, not only scalar figures of merit, and that generalize across device families and wavelengths.

\paragraph{Physics-informed and physics-constrained training.}
Physics-informed neural networks (PINNs) offer a general way to enforce PDE residuals in the loss for forward and inverse problems~\cite{Raissi2019PINNs}; in photonics this often means wave-equation residuals, such as Helmholtz, or boundary and energy constraints. However, PINNs typically optimize per-instance and do not amortize across a family of geometries. Neural operators such as the Fourier Neural Operator~\cite{Li2021FNO} and DeepONet~\cite{Lu2021DeepONet} learn mappings between function spaces and have been applied to PDE problems at scale, but the photonic setting still requires accurate amplitude and phase prediction because interference effects determine functionality. Our approach uses a Helmholtz residual penalty on predicted complex fields as a lightweight, architecture-agnostic physics constraint that amortizes across thousands of geometries.

\paragraph{Generative models for PDE-governed systems.}
Generative models can represent distributions over fields and enable conditional sampling for design. Score-based diffusion models~\cite{Song2021ScoreSDE} have been applied to PDE-governed problems: Huang \emph{et al.}\ use diffusion-based generation to solve Helmholtz and other equations under partial observation~\cite{Huang2024DiffusionPDE}. Flow matching~\cite{Lipman2023FlowMatching} provides a simulation-free alternative by learning a time-dependent velocity field along a probability path, offering straighter transport paths and simpler training than diffusion. Baldan \emph{et al.}\ extend this idea to physics-constrained generation with physics-based flow matching (PBFM), integrating PDE residuals during training~\cite{Baldan2025PBFM}. Our work is closest to PBFM, but targets electromagnetic fields and silicon photonic device families, combining a U-Net field generator with a Helmholtz residual constraint and phase-aware supervision.

\paragraph{Inverse design.}
Inverse design in photonics is often done with gradient-based optimization coupled to full-wave solvers~\cite{Molesky2018InverseDesign}. Classical adjoint-based methods have enabled compact devices such as broadband wavelength demultiplexers~\cite{Piggott2015DemultiplexerNatPhoton} and have been extended with fabrication constraints~\cite{Piggott2017FabricationConstrainedSciRep}; DeepAdjoint integrates data-driven models with optimization algorithms~\cite{Yeung2023DeepAdjoint}, and recent reviews summarize how deep learning and adjoint methods can be combined~\cite{Pan2023PhotonicsReview}. In contrast, we focus on the forward prediction problem of replacing the repeated field-solve step with a physics-constrained field generator, a prerequisite for any differentiable surrogate-based inverse design workflow. Additionally, our framework can be applied to train a generative conditional flow matching model conditioned on target S-parameters analogous to setting a cost function in inverse design, although this is outside the scope of this work and left for future exploration.

\section{\label{sec:level3}Background}
\subsection{Electromagnetic simulation of photonic devices}
We consider two-dimensional TE-polarized electromagnetic fields in silicon-on-insulator (SOI) waveguides. The out-of-plane electric field component $E_z(x,y)$ satisfies the scalar Helmholtz equation in the frequency domain:
\begin{equation}
    \nabla^2 E_z(x,y) + k_0^2 \,\varepsilon(x,y)\, E_z(x,y) = 0,
    \label{eq:helmholtz}
\end{equation}
\\
where $k_0=2\pi/\lambda$ is the free-space wavenumber and $\varepsilon(x,y)$ is the spatially varying relative permittivity. The reduction from three dimensions to two in this work is valid when the vertical slab mode has been pre-solved: for a 220\,nm SOI slab the fundamental TE mode yields an effective core index $n_\text{eff}\approx 2.4$
($\varepsilon_\text{core}\approx 5.8$) and the silica cladding has $n_\text{clad}=1.444$ ($\varepsilon_\text{clad}\approx 2.09$)~\cite{ReedKnights2004SiliconPhotonics}. We operate in the telecommunications C-band with a nominal wavelength of
$\lambda = 1.55~\mu$m.

Finite-difference time-domain (FDTD) methods solve the time-domain Maxwell's equations on a Yee grid, exciting the structure with an eigenmode source and extracting frequency-domain fields via discrete Fourier transforms~\cite{TafloveHagness2005FDTD}. Each simulation runs until field energy decays below a convergence threshold, typically requiring seconds to minutes for the domain sizes and resolution considered in this work. The key outputs used are the complex field $E_z(x,y)$, the permittivity map $\varepsilon(x,y)$, and source/port masks that identify the device excitation and measurement regions. Perfectly matched layers (PML) are used for the boundary conditions on all sides of the simulation domain.

\begin{figure*}[!ht]
\centering
\includegraphics[width=\textwidth]{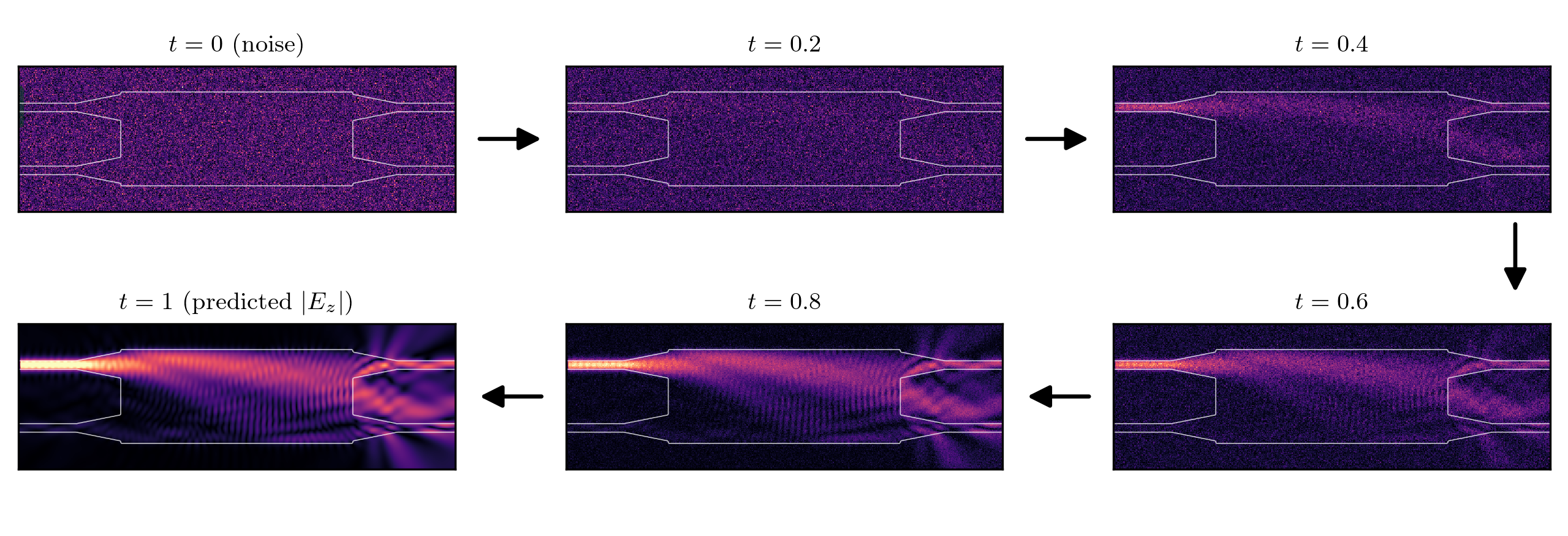}
\caption{Flow-matching transport visualized on a held-out $2\times 2$ MMI. Starting from a sample of Gaussian noise $x_0$ at $t=0$ (top left, with the active input port marked in green), the trained velocity field $u_\theta(x_t, t)$ is integrated forward in $t$ to produce the predicted complex field $E_z$ at $t=1$ (bottom left). The midpoint $t=0.5$ shows the field structure already emerging from noise. Panels share the same color scale; the integration uses 100 Euler steps.}
\label{fig:flowmatching_panel}
\end{figure*}

\subsection{Conditional Flow Matching}
Flow matching~\cite{Lipman2023FlowMatching, Liu2023RectifiedFlow} is a generative modeling framework that learns a time-dependent velocity field $u_\theta(x_t,t)$ to transport samples from a simple prior $p_0$ (Gaussian noise) to the data distribution $p_1$ along a deterministic path. Given source-target pairs $x_0\sim\mathcal{N}(0,I)$ and $x_1\sim p_\text{data}$, the conditional interpolation path is
\begin{equation}
    x_t = (1-t)\,x_0 + t\,x_1, \quad t\in[0,1],
    \label{eq:interpolation}
\end{equation}
with corresponding target velocity $u_t=x_1-x_0$. A neural network $u_{\theta}$ is trained to match this velocity by minimizing the loss function
\begin{equation}
    \mathcal{L}_\text{FM} = \mathbb{E}_{t,\,x_0,\,x_1}\!\left[\left\|u_\theta(x_t,t) - u_t\right\|^2\right].
    \label{eq:fm_loss}
\end{equation}
At inference, new samples are generated by integrating the learned velocity field from $t=0$ to $t=1$ using an ODE solver, starting from Gaussian noise. Compared with diffusion models, flow matching yields straighter transport paths that require fewer integration steps and avoids the need to tune a noise schedule~\cite{Lipman2023FlowMatching,Liu2023RectifiedFlow}.

In our setting, the data distribution $p_1$ consists of physically valid electromagnetic field solutions conditioned on device geometry and wavelength. The model learns to denoise a random field into one that satisfies the Helmholtz equation for a given permittivity map, a task we augment with explicit physics constraints, described in Section ~\ref{sec:method}. Figure~\ref{fig:flowmatching_panel} illustrates this transport on a held-out MMI: the noise is steadily reorganized into the predicted $|E_z|$ as $t$ advances along the ODE.
\section{\label{sec:method}Model Architecture}

\begin{figure*}[!t]
\centering
\begin{tikzpicture}[
    >=Stealth,
    every node/.style={font=\small},
    inputbox/.style={draw, rounded corners=2pt, fill=black!4,
        minimum width=72pt, minimum height=24pt, align=center, font=\footnotesize},
    noisebox/.style={draw, rounded corners=2pt, fill=red!5,
        minimum width=72pt, minimum height=24pt, align=center, font=\footnotesize},
    backbonebox/.style={draw, dashed, thick, rounded corners=3pt, fill=blue!6,
        minimum width=140pt, minimum height=66pt, align=center},
    odebox/.style={draw, rounded corners=2pt, fill=red!5,
        minimum width=140pt, minimum height=52pt, align=center, font=\footnotesize},
    outbox/.style={draw, rounded corners=2pt, fill=green!10,
        minimum width=84pt, minimum height=52pt, align=center, font=\footnotesize},
    arrow/.style={->, thick, >=Stealth}
]
\node[inputbox] (eps) at (0, 1.65)  {$\varepsilon(x,y)$\\\scriptsize permittivity};
\node[inputbox] (src) at (0, 0.55)  {$m_\mathrm{src}(x,y)$\\\scriptsize source mask};
\node[inputbox] (wl)  at (0, -0.55) {$\lambda$\\\scriptsize wavelength};
\node[noisebox] (noise) at (0, -2.1) {$x_0 \sim \mathcal{N}(0,I)$\\\scriptsize Gaussian noise};

\node[backbonebox] (backbone) at (5.5, 0.55) {%
    \textbf{Velocity-field predictor} $u_\theta$\\[2pt]
    \scriptsize $u_\theta(x_t,\, t \mid \varepsilon,\, m_\mathrm{src},\, \lambda)$\\[3pt]
    \scriptsize \textit{This work:} real-valued U-Net (63.3\,M params)\\
    \scriptsize \textit{Swappable:} DiT, FNO, \dots
};

\node[odebox] (ode) at (5.5, -2.1) {%
    \textbf{ODE integrator} (iterate $K$ times)\\
    Euler:\; $x_{k+1} = x_k + \Delta t_k \, u_\theta(x_k, t_k)$\\
    or 2nd-order Heun
};

\node[outbox] (out) at (11, -0.5) {%
    \textbf{Predicted}\\$E_z(x,y) \in \mathbb{C}$\\
    \scriptsize Re, Im channels
};

\draw[arrow] (eps.east) -- ([yshift=18pt]backbone.west);
\draw[arrow] (src.east) -- (backbone.west);
\draw[arrow] (wl.east)  -- ([yshift=-18pt]backbone.west);
\draw[arrow] (backbone.south) -- (ode.north) node[midway, right=2pt] {\scriptsize $u_\theta$};
\draw[arrow] (noise.east) -- (ode.west);
\draw[arrow] (ode.east) -- (out.west);
\end{tikzpicture}
\caption{System-level architecture of PIC-Flow. Conditioning $(\varepsilon, m_\mathrm{src}, \lambda)$ and a Gaussian noise sample $x_0$ enter a flow-matching pipeline whose backbone $u_\theta$ predicts a velocity field on the joint space of (current state, time, conditioning). The ODE integrator advances $x_t$ from $t=0$ to $t=1$ over $K$ steps, evaluating $u_\theta$ once per Euler step (or twice per Heun step), to produce the predicted complex field $E_z$. The dashed boundary on the velocity-field predictor indicates that the backbone is architecturally swappable: any module that regresses a 2D velocity field given the same conditioning could be substituted without changing the surrounding flow-matching pipeline, the physics losses, or the compliance metric.}
\label{fig:architecture}
\end{figure*}
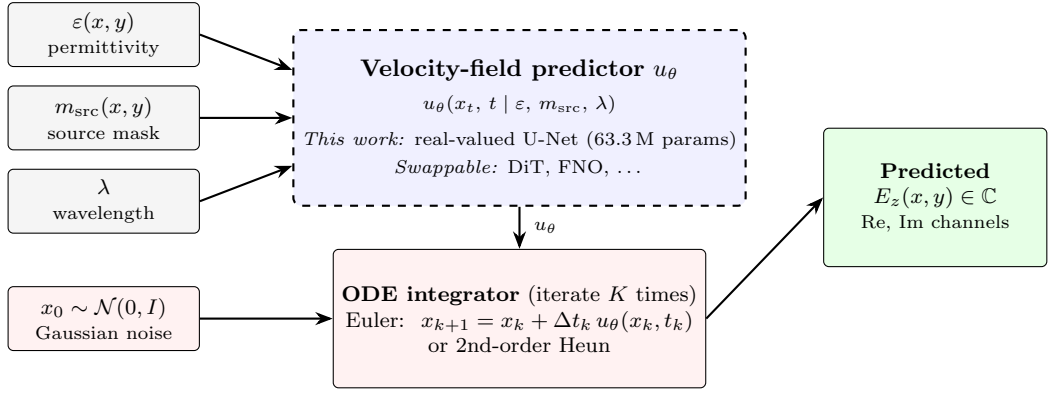

\subsection{Problem formulation}
Given a device permittivity map $\varepsilon(x,y)$, a source excitation mask $m_\text{src}(x,y)$, and a free-space wavelength $\lambda$, we seek the complex electric field $E_z(x,y)\in\mathbb{C}$ that solves the Helmholtz equation [Eq.~\eqref{eq:helmholtz}]. We frame this as a conditional generation problem: learn a model that maps Gaussian noise to the field solution, conditioned on $(\varepsilon, m_\text{src}, \lambda)$. Figure~\ref{fig:architecture} sketches the full pipeline. The model input is a tensor $\mathbf{x}\in\mathbb{R}^{4\times H\times W}$ whose channels are the real and imaginary parts of $E_z$ (interpolated between noise and target during training), the permittivity map, and the source mask. The wavelength enters as a scalar conditioning variable through the timestep embedding. The model outputs a two-channel velocity field $\hat{u}_\theta\in\mathbb{R}^{2\times H\times W}$ representing the real and imaginary components of the predicted flow-matching velocity.

\subsection{U-Net architecture}

We use a real-valued U-Net with 63.3\,M parameters that treats the real and imaginary parts of $E_z$ as separate channels. The architecture follows a standard encoder-bottleneck-decoder layout with residual convolutional blocks at each level, multi-head self-attention at downsample factors 4 and 8, and conditioning on the flow-matching timestep $t$ and normalized wavelength $\lambda$ via Feature-wise Linear Modulation (FiLM), in which the conditioning variables are passed through a small MLP to produce per-layer scale and shift parameters that modulate the network's intermediate features, allowing the model to adapt its behavior at each layer based on the current timestep and wavelength. The permittivity map and source mask are concatenated with the field channels at the input so the model has direct access to material layout and excitation location at every depth. Block sizes, attention layout, and conditioning details can be found in Appendix ~\ref{app:arch}.

\subsection{Training objective}

The ablation runs in this work use a weighted subset of three losses:
\begin{equation}
    \mathcal{L} = \mathcal{L}_\text{FM}
    + \lambda_R\,\mathcal{L}_\text{res}
    + \lambda_\phi\,\mathcal{L}_\text{phase},
    \label{eq:total_loss}
\end{equation}
where $\mathcal{L}$ is the total loss, $\mathcal{L_\text{FM}}$ is the flow matching loss, $\mathcal{L}_\text{res}$ is the Helmholtz residual loss, and $\mathcal{L}_\text{phase}$ is the phase loss, with weights $\lambda_R$ and $\lambda_\phi$ set by the experiment design in Section~\ref{sec:level6}. Other auxiliary losses are disabled for these three tests.

\paragraph{Flow matching loss.}
Given source--target pairs $(x_0, x_1)$ where $x_0 \sim \mathcal{N}(0,I)$ and $x_1$ is the ground-truth field, we sample time $t$ from a 50/50 mixture of $\mathcal{U}[0,1]$ and a logit-normal distribution $t = \sigma(\mu + s\cdot z)$, $z \sim \mathcal{N}(0,1)$. In these experiments $\mu=0.8$ and $s=0.8$, so the nonuniform half of the sampler emphasizes intermediate-to-late denoising states while the uniform half preserves endpoint coverage. The interpolated state is $x_t = (1-t)x_0 + t\,x_1$ and the target velocity is $u_t = x_1 - x_0$. The loss is the spatially weighted MSE between the predicted and target velocities:
\begin{equation}
    \mathcal{L}_\text{FM} = \frac{\sum_{p} w(p)\,\|\hat{u}_\theta(x_t,t)(p) - u_t(p)\|^2}{\sum_{p} w(p)},
    \label{eq:fm_loss_weighted}
\end{equation}
where the spatial weight $w(p)$ combines a PML-exclusion mask with an optional device-region upweight.

\paragraph{Helmholtz residual loss.}
From the predicted velocity we reconstruct the clean field estimate
$\hat{x}_1 = x_t + (1-t)\hat{u}_\theta$ and de-normalize to physical units. The Helmholtz residual is
\begin{equation}
    R(p) = \nabla^2 \hat{E}_z(p) + k_0^2\,\varepsilon(p)\,\hat{E}_z(p),
    \label{eq:residual}
\end{equation}
computed with a five-point finite-difference Laplacian. The loss is the masked, normalized mean-squared residual:
\begin{equation}
    \mathcal{L}_\text{res} = \frac{\sum_p\, w_R(p)\,|R(p)|^2}{\sum_p\, w_R(p)\,|k_0^2\varepsilon(p) E_z^\text{true}(p)|^2},
    \label{eq:residual_loss}
\end{equation}
where the weight mask $w_R$ is the product of three binary masks: (i) a PML-exclusion mask that zeros out absorbing boundary cells, (ii) a source-exclusion mask (the Helmholtz equation is only valid in source-free regions), and (iii) an interface mask that excludes pixels where $|\nabla\varepsilon|$ exceeds a threshold, since the five-point Laplacian stencil produces spurious residuals at sharp dielectric boundaries (Si/SiO$_2$). The denominator normalizes by the driving-term scale $|k_0^2\varepsilon E_z|^2$, making the loss approximately 0 when physics is satisfied and 1 when the residual equals the driving term, independent of grid spacing and field amplitude.

For per-sample physics-fidelity reporting we further define the dimensionless compliance percentage
\begin{equation}
    \rho_R \;=\; \sqrt{\mathcal{L}_\text{res}}\,\times\,100\%,
    \label{eq:compliance}
\end{equation}
which has a direct physical interpretation as the RMS Helmholtz residual relative to the local driving-term magnitude $|k_0^2\,\varepsilon\,E_z|$. A field that exactly satisfies the wave equation gives $\rho_R = 0$; FDTD's own discretization residual on this dataset sits below $0.1\%$. We use $\rho_R$ to annotate the qualitative comparisons in Figs.~\ref{fig:inference_examples} and~\ref{fig:ood_inference_examples} and report numerical compliance in Table~\ref{tab:compliance}.

When time-gating is enabled, samples with $t$ below a threshold are excluded from the residual computation, since the reconstructed field at low $t$ is too noisy to yield informative physics gradients.

\paragraph{Phase loss.}
We penalize phase disagreement between the predicted and ground-truth fields using a wrap-safe cosine distance:
\begin{equation}
    \mathcal{L}_\text{phase} = 1 - \mathrm{Re}\!\left(\frac{\hat{E}_z}{|\hat{E}_z|} \cdot
    \frac{\bar{E}_z^\text{true}}{|E_z^\text{true}|}\right),
    \label{eq:phase_loss}
\end{equation}
weighted by $|E_z^\text{true}|^2$ and gated by an amplitude threshold so that phase is only supervised in regions with non-negligible field strength. Before computing the phase loss, predicted and true fields are globally phase-aligned via a weighted inner product to remove the gauge freedom inherent in complex field solutions.

\subsection{\label{sec:curriculum}Inference and loss schedules}

At inference, we generate field predictions by integrating the learned velocity field from $t=0$ (Gaussian noise) to $t=1$ on a linear time grid using either an explicit Euler step or the second-order Heun integrator (Appendix~\ref{app:integrators}). The wall-clock benchmark in Section~\ref{sec:walltime} uses Euler; the qualitative figures use Heun. Across the three ablation runs the FM loss is active at every epoch, the phase loss ramps from epoch~1 to $\lambda_\phi=0.1$ over 15 epochs when enabled, and the Helmholtz residual loss ramps in from epoch~11 to $\lambda_R=1.0$ over 20 epochs when enabled.

\section{\label{sec:level5}Dataset Generation}

\subsection{Device Types}
The ablation study in Section~\ref{sec:level6} uses a three-family silicon photonics dataset at a single C-band wavelength, $\lambda=1.55\,\mu$m. The device families span interference, splitting, and evanescent coupling, and all runs use the same train/validation split and sample-evaluation protocol. All devices use silicon-on-insulator waveguides with an effective core index $n_\text{eff}\approx2.4$ (from pre-computed eigenmode tables for a 220\,nm slab) and SiO$_2$ cladding with $n_\text{clad}=1.444$. In the normalized simulation tensors, these correspond to $\varepsilon_\text{core}=5.8$ and $\varepsilon_\text{clad}=2.09$. The simulation grid is fixed at $8~\mu$m $\times$ $24~\mu$m with 50 nm resolution, corresponding to 31 points per wavelength at $\lambda=1.55\,\mu$m. The ablation runs exclude straight waveguides, tapers, bends, S-bends, and crossings which exist in the codebase so that the comparison focuses on the three more interesting high-dimensional device families in Table~\ref{tab:devices}.

\begin{table}[!htbp]
\centering
\caption{Device families and their geometry sweep ranges. Each family is sampled in a 5-dimensional parameter space using Latin hypercube sampling, with values quantized to a half-pixel grid (0.025\,$\mu$m at 20 pixels/$\mu$m) so that each parameter change produces a measurable change in the permittivity map. \emph{Count} is the number of simulated geometries at $\lambda=1.55\,\mu$m.}
\label{tab:devices}
\begin{ruledtabular}
\begin{tabular}{llc}
Family / Parameter & Range ($\mu$m) & Count \\
\hline
\textbf{2$\times$2 MMI} (4 ports)        &              & 7,500 \\
\quad waveguide width                    & 0.40--0.575  &       \\
\quad MMI width                          & 4.5--5.5     &       \\
\quad MMI length                         & 8.0--15.0    &       \\
\quad taper width                        & 0.575--1.5   &       \\
\quad taper length                       & 1.0--3.0     &       \\[2pt]
\textbf{Y-branch} (3 ports)              &              & 7,500 \\
\quad waveguide width                    & 0.40--0.575  &       \\
\quad junction length                    & 1.0--3.0     &       \\
\quad bend length                        & 4.0--7.0     &       \\
\quad arm offset                         & 0.575--2.5   &       \\
\quad output length                      & 1.0--4.0     &       \\[2pt]
\textbf{Directional coupler} (4 ports)   &              & 7,500 \\
\quad waveguide width                    & 0.40--0.575  &       \\
\quad gap $g$                            & 0.10--0.35   &       \\
\quad coupling length $L_c$              & 5.0--8.0     &       \\
\quad bend length                        & 4.0--6.0     &       \\
\quad port separation                    & 0.825--2.0   &       \\[2pt]
\hline
Total                                    &              & 22,500 \\
\end{tabular}
\end{ruledtabular}
\end{table}

\subsection{FDTD simulation and splits}
Each geometry is simulated with the Meep FDTD solver~\cite{Oskooi2010Meep} at $\lambda=1.55\,\mu$m and stored as a $160\times480$ pixel tensor of $\mathrm{Re}(E_z)$, $\mathrm{Im}(E_z)$, normalized permittivity, and source mask, plus per-port binary masks. The 22{,}500 samples are split 18{,}000 / 2{,}250 / 2{,}250 into train, validation, and held-out test, with all three ablation runs sharing the same index-based split so differences in metrics reflect the loss configuration rather than different data partitions. Frequency-domain fields have an arbitrary global phase; we anchor each FDTD target to its source-region phase to remove this gauge freedom before training. Eigenmode source configuration, exact pixel resolution, and the phase-anchoring algorithm are described in detail in Appendix~\ref{app:fdtd}--\ref{app:phase_anchor}.

\begin{figure*}[!t]
\centering
\includegraphics[width=\textwidth]{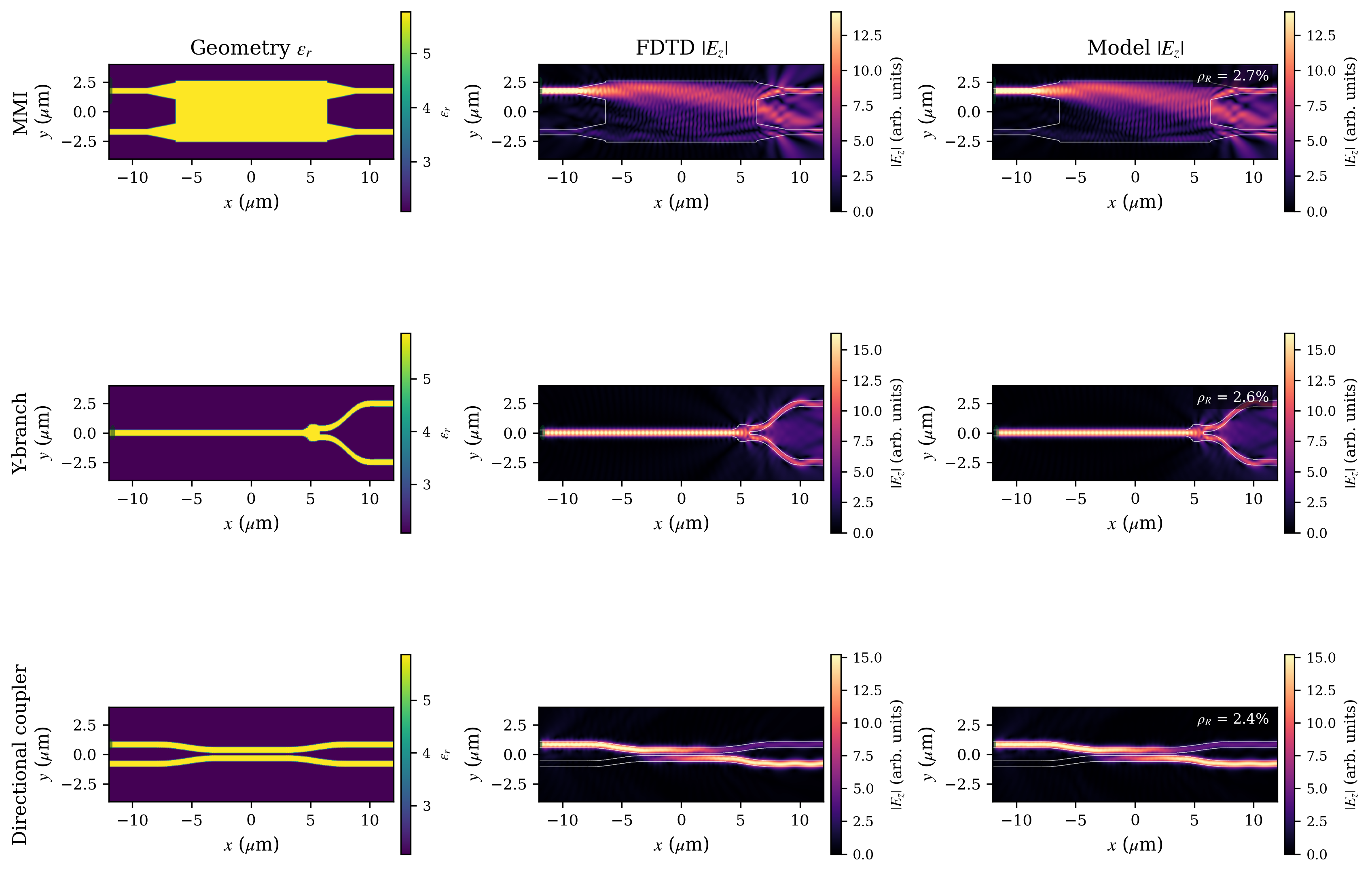}
\caption{Held-out qualitative comparison for parameterized devices drawn from the training distribution. Each row is a random \texttt{test}-split sample for an MMI (top), Y-branch (middle), and directional coupler (bottom). Columns show (left) $\varepsilon$ with the sourced input port highlighted in green, (middle) FDTD $|E_z|$, and (right) model $|E_z|$. Sampling uses the epoch-300 FM+phase+residual checkpoint with $200$ Heun steps and the checkpoint-trained time grid; fields are shown in arbitrary units with per-panel color scaling. Each Model panel is annotated with the per-sample compliance percentage $\rho_R$ defined in Eq.~\eqref{eq:compliance}.}
\label{fig:inference_examples}
\end{figure*}

\section{\label{sec:level6}Experiments}

We compare three matched real-valued U-Net runs. All use the same architecture (63.3M parameters), the same $160\times480$ field grid, the same training/validation split, the same $100$-step linear flow sampler for sample evaluation, and the same fixed stratified sample-evaluation set of 24 devices (8 per family). All three runs completed for 300 epochs on 12 NVIDIA V100 GPUs with global batch size 48. Sampling settings differ by purpose: training-time evaluation and the ablation metrics in Table~\ref{tab:ablation_final_metrics} use a 100-step Euler sampler, the qualitative comparison figures use 200-step Heun for the cleanest visualization, and the wall-clock benchmark in Section~\ref{sec:walltime} sweeps Euler step counts to characterize the speed-versus-compliance tradeoff.

\begin{table*}[!tbp]
\centering
\caption{Loss configurations for the ablation study. Phase loss, when enabled, ramps from epoch 1 to $\lambda_\phi=0.1$ over 15 epochs. Residual loss, when enabled, starts at epoch 11 and ramps to $\lambda_R=1.0$ over 20 epochs.}
\label{tab:ablation_design}
\small
\setlength{\tabcolsep}{5pt}
\begin{tabular}{lcccccc}
\hline\hline
Run & Objective & GPUs & Global batch & $\lambda_R$ & $\lambda_\phi$ \\
\hline
FM only & $\mathcal{L}_\mathrm{FM}$ & 12$\times$V100 & 48 & 0 & 0 \\
FM + phase & $\mathcal{L}_\mathrm{FM}+\lambda_\phi\mathcal{L}_\mathrm{phase}$ & 12$\times$V100 & 48 & 0 & 0.1 \\
FM + phase + residual & $\mathcal{L}_\mathrm{FM}+\lambda_\phi\mathcal{L}_\mathrm{phase}+\lambda_R\mathcal{L}_\mathrm{res}$ & 12$\times$V100 & 48 & 1.0 & 0.1 \\
\hline\hline
\end{tabular}
\end{table*}

\begin{table*}[!tbp]
\centering
\caption{Final validation/sample-evaluation metrics at epoch 300. Residual ratio is the mean sample Helmholtz residual divided by the mean ground-truth residual on the fixed 24-sample evaluation set; lower is better. PSNR is computed from sampled fields.}
\label{tab:ablation_final_metrics}
\footnotesize
\setlength{\tabcolsep}{5pt}
\begin{tabular}{lccccccc}
\hline\hline
Run & Epoch & Val. FM & Val. residual & Val. phase & Residual ratio & Worst p95 ratio & PSNR (dB) \\
\hline
FM only & 300 & 0.00668 & -- & -- & 1.546 & 2.553 & 37.92 \\
FM + phase & 300 & 0.00664 & -- & 0.000662 & 1.531 & 2.502 & 37.97 \\
FM + phase + residual & 300 & 0.00787 & 0.00245 & 0.000801 & 1.336 & 2.222 & 37.41 \\
\hline\hline
\end{tabular}
\end{table*}

The ``Residual ratio'' column in Table~\ref{tab:ablation_final_metrics} reports the per-sample RMS predicted Helmholtz residual divided by the corresponding FDTD residual on the fixed evaluation set. Because the driving-term scale cancels in the ratio, this quantity is mathematically equal to $\rho_R^\mathrm{pred}/\rho_R^\mathrm{FDTD}$ and therefore captures the same physics as the compliance percentage in Eq.~\eqref{eq:compliance}, expressed multiplicatively against the FDTD discretization floor. We retain the multiplicative form here for run-to-run comparison and use $\rho_R$ as the per-sample reporting form (Table~\ref{tab:compliance}); the latter has the advantage of a direct interpretation as residual-relative-to-driving-force and does not depend on the FDTD noise floor.

The completed runs show a consistent tradeoff. Phase supervision alone gives a small improvement in final worst-device residual ratio relative to FM-only (2.50 versus 2.55 at epoch 300), while adding the Helmholtz residual yields the strongest physics-compliance improvement (2.22 final worst p95 ratio and 1.34 mean residual ratio). The residual term slightly reduces final PSNR relative to FM+phase, suggesting that the PDE penalty shifts optimization toward wave-equation compliance rather than purely pixel-level reconstruction.

\subsection{Qualitative inference examples}

Beyond the aggregate metrics in Table~\ref{tab:ablation_final_metrics}, Figure~\ref{fig:inference_examples} shows the FM+phase+residual checkpoint on three random held-out \texttt{test} samples (one MMI, one Y-branch, and one directional coupler), and Table~\ref{tab:compliance} reports the per-sample compliance $\rho_R$ (Eq.~\eqref{eq:compliance}) for these and for the out-of-distribution (OOD) geometries in Section~\ref{sec:ood_examples}. In-distribution compliance sits at $2.2$--$2.7\%$ across families, well above the FDTD discretization floor ($<0.1\%$) but small enough that the predicted $|E_z|$ is almost visually indistinguishable from the FDTD reference at the resolution of the figure.

\begin{table}[!htbp]
\centering
\caption{Per-sample compliance percentage $\rho_R$ (Eq.~\eqref{eq:compliance}) for the qualitative samples in Fig.~\ref{fig:inference_examples} (in-distribution) and Fig.~\ref{fig:ood_inference_examples} (out-of-distribution, Appendix~\ref{sec:ood_examples}). Lower is better; FDTD's own discretization residual on this dataset sits below $0.1\%$.}
\label{tab:compliance}
\begin{ruledtabular}
\begin{tabular}{lc}
Sample & $\rho_R$ \\
\hline
\multicolumn{2}{l}{\textit{In-distribution (Fig.~\ref{fig:inference_examples})}}\\
$2\times2$ MMI                  & $2.7\%$ \\
Y-branch                        & $2.5\%$ \\
Directional coupler             & $2.2\%$ \\[2pt]
\multicolumn{2}{l}{\textit{Out-of-distribution (Fig.~\ref{fig:ood_inference_examples})}}\\
S-bend (tight $R$, large offset) & $12\%$ \\
Short, steep taper               & $4.0\%$ \\
Long, wide taper                 & $3.6\%$ \\
Cascaded $1\times3$ Y-branch     & $9.1\%$ \\
\end{tabular}
\end{ruledtabular}
\end{table}

\subsection{\label{sec:ood_examples}Out-of-distribution generalization}

\begin{figure*}[!tbp]
\centering
\includegraphics[width=\textwidth]{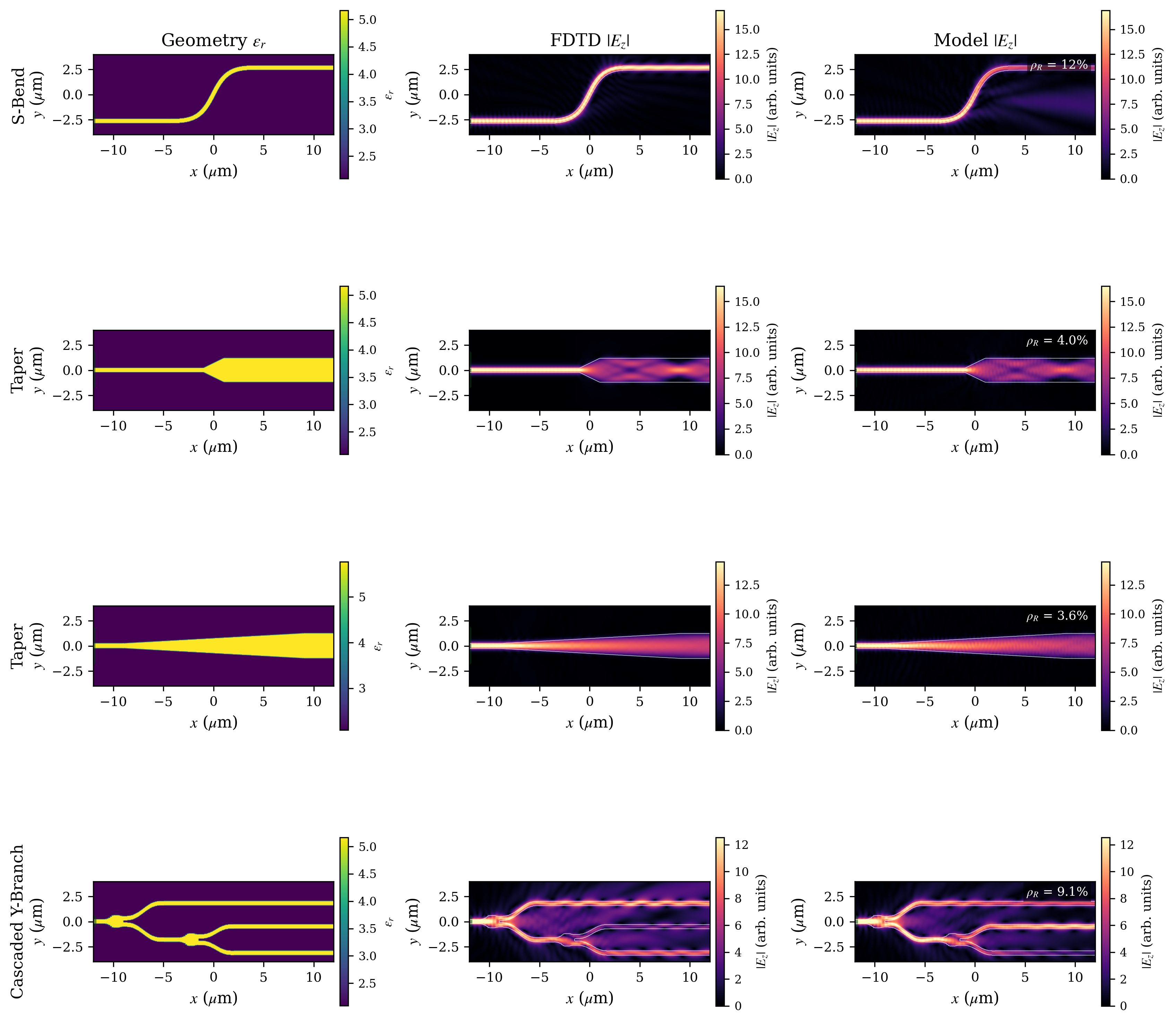}
\caption{Out-of-distribution qualitative comparison. Row 1: aggressive Euler S-bend ($R_\mathrm{min}=2.5\,\mu$m, lateral offset $6\,\mu$m). Row 2: short, steep taper ($w_\mathrm{out}=2.4\,\mu$m, length $2\,\mu$m). Row 3: long, wide taper ($w_\mathrm{out}=2.5\,\mu$m, length $18\,\mu$m). Row 4: cascaded $1\times3$ Y-branch built from two stacked Y-junctions, an unseen device class. Columns and sampling protocol match Fig.~\ref{fig:inference_examples}, and each Model panel reports the per-sample compliance percentage $\rho_R$.}
\label{fig:ood_inference_examples}
\end{figure*}

To probe behavior outside the training distribution we apply the same epoch-300 FM+phase+residual checkpoint to four held-out test cases: an aggressive Euler S-bend with bend radius below the trained sweep ($R_\mathrm{min}=2.5\,\mu$m vs.\ trained 3 -- 7 $\mu$m) and lateral offset above the trained range ($6\,\mu$m vs.\ 2 -- 5.5 $\mu$m); a short and steep taper that compresses the trained taper-length range; a long, wide taper with input--output widths and length all outside the trained taper sweep; and a cascaded $1\times3$ Y-branch formed from two stacked Y-junctions, an entirely new device class never seen during training. FDTD reference simulations are produced with the same Meep configuration to serve as ground-truth values for comparison. Figure~\ref{fig:ood_inference_examples} shows the FDTD reference and model prediction for these four cases; per-sample $\rho_R$ values appear in the lower rows of Table~\ref{tab:compliance}.

The model retains physically meaningful predictions on the parameter-OOD tapers (long-wide at $3.6\%$ and short-steep at $4.0\%$, both within roughly $1$--$2$ percentage points of the in-distribution families) and on a structurally new device class (the cascaded $1\times3$ Y-branch at $9.1\%$). The aggressive S-bend reaches $\rho_R \approx 12\%$, indicating a moderate but recognizable physics violation, namely that the field amplitude pattern is qualitatively correct but the wave equation is no longer satisfied at the trained-distribution level. The two tapers bracket the regime in which the model degrades gracefully as geometry parameters push further from the training sweep, rather than failing catastrophically.

\subsection{\label{sec:walltime}Wall-clock benchmark}

To quantify the runtime advantage of the surrogate over conventional FDTD, we benchmark inference wall time against Meep~\cite{Oskooi2010Meep} on a single held-out directional coupler. Both methods run on the same compute node (NVIDIA~A100 GPU, dual AMD~EPYC~75F3 host); FDTD uses Meep on 16 threads and the surrogate uses fp16 autocast with the FM+phase+residual checkpoint from Table~\ref{tab:ablation_final_metrics}. Detailed hardware, software, and sampler configuration can be found in Appendix ~\ref{app:walltime}.

\begin{table}[!htbp]
\centering
\caption{Wall-clock inference time and physics compliance ($\rho_R$) for the same held-out test-set directional coupler. FDTD reference uses Meep with 16 threads on dual AMD EPYC~75F3. The surrogate runs on a single NVIDIA~A100 with fp16 autocast, compiled kernels, and a channels-last memory layout. Speedup is over the FDTD reference on the same node. All entries are single-seed measurements; multi-seed variance is left for future work.}
\label{tab:walltime}
\begin{ruledtabular}
\begin{tabular}{lccc}
Method & Wall time & Speedup & $\rho_R$ \\
\hline
FDTD (Meep)                  & $5.61$\,s   & $1.0\times$    & (ref) \\
PIC-Flow, Euler 200-step     & $4.35$\,s   & $1.3\times$    & $1.2\%$ \\
PIC-Flow, Euler 100-step     & $2.19$\,s   & $2.6\times$    & $1.9\%$ \\
PIC-Flow, Euler 50-step      & $1.10$\,s   & $5.1\times$    & $2.1\%$ \\
PIC-Flow, Euler 20-step      & $440$\,ms   & $12.7\times$   & $3.0\%$ \\
PIC-Flow, Euler 10-step      & $220$\,ms   & $25.4\times$   & $3.9\%$ \\
PIC-Flow, Euler 5-step       & $110$\,ms   & $50.6\times$   & $5.5\%$ \\
PIC-Flow, Euler 1-step       & $22$\,ms    & $252\times$    & $14.5\%$ \\
\end{tabular}
\end{ruledtabular}
\end{table}

At 100 sampler steps the surrogate is $2.6\times$ faster than 16-thread CPU FDTD on the same node while reaching $\rho_R=1.9\%$, comparable to the in-distribution compliance levels reported in Table~\ref{tab:compliance}. Reducing the sampler to 20 steps drops the wall time to 440\,ms, a $12.7\times$ speedup, with $\rho_R=3.0\%$, still well below the $\sim\!12\%$ regime observed for heavily out-of-distribution geometries (Section~\ref{sec:ood_examples}). The compliance-versus-speed tradeoff is gradual through 5 steps (a $50.6\times$ speedup with $\rho_R=5.5\%$) but breaks down at 1 step, where $\rho_R$ jumps to $14.5\%$, comparable to the worst OOD geometry reported in this paper. Because the hardware is held fixed across both methods, this comparison isolates the surrogate-vs-FDTD speedup from any GPU-vs-CPU asymmetry. In practice, the designer can determine the optimal fidelity-to-speed tradeoff for their specific application and set the number of sampler steps accordingly.

\section{\label{sec:level7}Discussion}

The ablation results support the use of physics-informed supervision as a complement to flow matching rather than a replacement for the generative objective. Flow matching alone learns high-quality field samples, but the Helmholtz residual term directly targets a failure mode that pixel metrics can miss, namely fields that look plausible while violating the governing wave equation. The completed FM+phase+residual run improves the worst-device residual ratios relative to FM-only and FM+phase baselines, indicating that the residual penalty improves physical consistency on the fixed sample-evaluation set.

At the same time, the results show a tradeoff between reconstruction fidelity and physics compliance. The residual-augmented model has slightly lower final PSNR than the FM+phase model, suggesting that the PDE term changes the optimization target rather than simply improving every metric simultaneously. This is expected for a finite-capacity model trained with multiple losses, as a field can be closer in pixelwise mean-square error while still having larger local Helmholtz residuals, especially near interfaces, sources, and low-amplitude regions.

\paragraph{Scaling with domain size, resolution, and dimensionality.}
A practical implication of these results is that the surrogate's runtime advantage scales favorably with simulation size, grid resolution, and dimensionality. FDTD cost grows quickly along all three axes: each doubling of the grid increases the work per time step in proportion to the number of new cells, and the Courant--Friedrichs--Lewy stability condition forces a matching reduction in the time step, so the total cost grows faster than cell count alone~\cite{TafloveHagness2005FDTD}. This accuracy versus compute tradeoff is what makes 3D FDTD the bottleneck for integrated photonic device optimization, often forcing practitioners to coarsen grids or restrict sweeps to a handful of design points~\cite{BurrFarjadpour2005FDTDAccuracy}. PIC-Flow inference, in contrast, is dominated by a single U-Net forward pass whose cost grows only with the number of pixels at fixed model size, with no time-stepping multiplier on top. Larger domains, finer grids, and the move from 2D scalar to 3D vectorial fields therefore all widen the surrogate's relative advantage. The $12.7\times$ speedup at 20 Euler steps reported in Section~\ref{sec:walltime} (or the $50.6\times$ at 5 steps, with a graceful compliance penalty) can be reasonably interpreted as a floor. The same comparison on circuit-scale, finer-resolution, or 3D vectorial domains where FDTD currently bottlenecks photonic design is expected to widen substantially. Scaling the surrogate to those regimes is primarily a question of training-data generation and architecture choice rather than per-inference compute.

\paragraph{Current limitations.}
The current study has several limitations, which we discuss here in part to help motivate future work. First, the ablation dataset is restricted to three device families at a single wavelength, so generalization to broader wavelength bands and unseen topologies remains qualitative. While the training data was only produced for a 220 nm thick silicon device layer, this process can be extended to other material platforms (e.g. thin film lithium niobate or silicon nitride) and other wavelengths (e.g. O-band or visible). Additionally, the model predicts only the complex $E_z$ field, so additional electromagnetic quantities are not directly available from the surrogate. Furthermore, the U-Net architecture is not exactly equivariant under device symmetries, so symmetric devices may still show asymmetric sampled responses unless symmetry is enforced through architecture, paired conditioning, or sampling constraints. Finally, inference still requires multi-step ODE integration, and the speed--compliance curve in Section ~\ref{sec:walltime} shows that the model needs roughly twenty Euler steps to reach in-distribution-grade compliance, with the cost scaling linearly with step count. A second training pass such as reflow~\cite{Liu2023RectifiedFlow} or distillation could in principle reduce this to one or two steps, at the cost of an additional training pipeline beyond the present scope.

\paragraph{Future work.}
Future work will extend the dataset to more device classes, wavelengths, and materials, and investigate S-parameter prediction or extraction from inferred fields, but rigorous modal decomposition may require additional field components or a dedicated surrogate trained directly for scattering amplitudes. Furthermore, exploration of using PIC-Flow inference to replace forward solve steps in inverse design optimization loops presents a potentially fruitful opportunity to accelerate the design cycle relative to current FDTD-bound standards. Reducing the number of integration steps is also a priority, and a second training stage such as reflow~\cite{Liu2023RectifiedFlow} or step-distillation could push the sampler toward one- or two-step generation, multiplying the wall-clock advantage in Section~\ref{sec:walltime} without sacrificing the physics-aware training signal.

\section{\label{sec:level8}Conclusion}

We proposed and demonstrated a real-valued U-Net flow-matching model for generating steady-state electromagnetic fields in parameterized silicon photonic devices. Our model completely reconstructs the spatial electric field profile of arbitrary photonic devices on an  $8 \times 24~\mu$m simulation domain with high fidelity. The model conditions on device geometry, source-port masks, and wavelength, and learns to transform Gaussian field noise into complex-valued $E_z$ solutions on a fixed spatial grid. Across matched ablations, adding Helmholtz residual supervision improves physics-compliance metrics relative to FM-only and FM+phase baselines, while preserving high sample fidelity.

These results suggest that flow matching provides a practical generative framework for fast photonic field prediction, and that explicit wave-equation penalties can improve the physical reliability of sampled fields compared to flow matching alone. While we do not envision PIC-Flow as a standalone FDTD replacement, the dramatic speedups while maintaining reasonable fidelity can enable rapid design space exploration of large parameter spaces, replacing brute-force multidimensional parameter sweeps. In this way, our model can be used for rapidly paring down a large design space and in turn reduce the total number of calls to costly FDTD solvers, which can be reserved for only final fine-tuning sweeps. Finally, our model, training data, and weights are all made fully open-source, enabling integrated photonics designers to experiment with different workflows incorporating PIC-Flow with existing tools, fine-tune weights for specific applications, add new training data cases, and explore other promising network architectures.

\section{\label{sec:level9}Data and Code Availability}

Code, model, and dataset are available at
\href{https://github.com/Rizzo-Integrated-Photonic-Systems-Lab/PIC-Flow}{\textcolor{blue}{GitHub}},
\href{https://huggingface.co/RizzoLab/PIC-Flow}{\textcolor{blue}{HuggingFace (model)}}, and
\href{https://huggingface.co/datasets/RizzoLab/PIC-Flow-Dataset}{\textcolor{blue}{HuggingFace (dataset)}}.

\section{\label{sec:level10}Acknowledgments}

We acknowledge funding from Dartmouth College under the Undergraduate Research Assistantships at Dartmouth (URAD) program and startup funding support from the Dartmouth College Thayer School of Engineering.

\appendix

\section{\label{app:impl}Implementation details}

\subsection{\label{app:arch}U-Net architecture}

The architecture follows the standard encoder--bottleneck--decoder layout with skip connections, residual blocks, and attention at selected spatial resolutions.

\paragraph{Residual convolutional blocks.}
Each resolution level contains convolutional residual blocks with normalization, nonlinear activation, and dropout. Downsampling in the encoder increases the channel count while reducing spatial resolution; upsampling in the decoder restores the original grid and combines decoder features with encoder skip connections.

\paragraph{Self-attention.}
At downsample factors 4 and 8 (corresponding to $40\times120$ and $20\times60$ token grids on the $160\times480$ field), we apply multi-head self-attention over spatial tokens. Attention lets features at one location directly reference features at distant locations, complementing the locality of convolutional layers.

\paragraph{Conditioning.}
The flow-matching timestep $t$ is encoded via sinusoidal positional embeddings and concatenated with the normalized wavelength $\lambda$. This combined embedding passes through a two-layer MLP to produce per-layer scale and shift parameters $(\gamma, \beta)$ that modulate each residual block's features via FiLM-style modulation (as described in Section~\ref{sec:method}), allowing the model to adapt its behavior at each layer based on the current timestep and wavelength.

\paragraph{Auxiliary inputs.}
The permittivity map and source mask are concatenated with the real and imaginary field channels before the U-Net stem. This gives the model direct access to material layout and excitation location throughout the prediction process.

\subsection{\label{app:integrators}Inference integrators}

We integrate the learned velocity field from $t=0$ (Gaussian noise) to $t=1$ on a linear time grid using one of two integrators. The first is the explicit (first-order) Euler step
\begin{equation}
    x_{k+1} = x_k + \Delta t_k\, \hat{u}_\theta(x_k, t_k),
    \label{eq:euler}
\end{equation}
which requires a single network evaluation per step and is what the wall-clock benchmark in Sec.~\ref{sec:walltime} uses. The second is the second-order Heun (improved Euler) integrator,
\begin{align}
    \tilde{x}_{k+1} &= x_k + \Delta t_k\, \hat{u}_\theta(x_k, t_k), \nonumber\\
    x_{k+1} &= x_k + \tfrac{1}{2}\Delta t_k\!\left[\hat{u}_\theta(x_k, t_k)
               + \hat{u}_\theta(\tilde{x}_{k+1}, t_{k+1})\right],
    \label{eq:heun}
\end{align}
which doubles the cost per step but reduces local truncation error. The ablation runs use a linear time grid with 100 flow-matching steps.

\subsection{\label{app:fdtd}FDTD simulation parameters}

Meep simulations use an eigenmode source exciting the fundamental TE mode at the selected input port, run until field energy decays below a convergence threshold (typically seconds to minutes per geometry on CPU). The training tensors are cropped/resampled to a $160\times480$ grid at $dx=dy=0.05\,\mu$m. Per simulation we store the complex field $E_z$, the permittivity map $\varepsilon$, a source mask identifying the excitation port, per-port binary masks at each port location, and the four-port S-parameter vector. Geometry parameters are stored alongside each sample and indexed in a shared shard manifest used by the dataloader.

\subsection{\label{app:phase_anchor}Phase anchoring}

Frequency-domain fields have an arbitrary global phase, meaning that if $E_z$ is a solution to the Helmholtz equation, so is $E_ze^{i\phi}$. Without phase anchoring, the same device at the same wavelength could produce training targets with different global phases, preventing the model from learning a consistent mapping.

We anchor the global phase using the field within the source region. A weighted average of unit phasors is computed in the source mask, weighted by local field intensity:
\begin{equation}
    \phi_\text{ref} = \arg\!\left(\sum_{p \in \mathcal{M}_\text{src}} |E_z(p)|^2 \cdot
    \frac{E_z(p)}{|E_z(p)|}\right),
    \label{eq:phase_anchor}
\end{equation}
and the entire field is rotated by $e^{-i\phi_\text{ref}}$, setting the intensity-weighted phase in the source region to approximately zero. When restricted to high-permittivity pixels ($\varepsilon > 3.0$, i.e.\ within the waveguide core), this anchoring is robust to the source mask geometry. If the source mask is unavailable or too small, we fall back to a fixed region-of-interest near the input waveguide. This two-stage strategy (mask-first, ROI fallback) ensures consistent targets even after D4 augmentation, which may rotate the source to a different edge of the domain.

\subsection{\label{app:walltime}Wall-clock benchmark configuration}

The wall-clock comparison in Section~\ref{sec:walltime} is performed on a single held-out directional coupler ($\mathrm{gap}=0.175\,\mu$m, $L_c=5.97\,\mu$m, $\lambda=1.55\,\mu$m). All measurements are taken on the same compute node: an NVIDIA~A100 GPU paired with dual AMD~EPYC~75F3 processors ($2\times32$ physical cores, Zen3). FDTD is run with Meep on 16 threads, in line with multi-threaded Meep configurations reported in recent photonic ML benchmarks. The surrogate runs PyTorch~2.x on the A100 with fp16 autocast, compiled kernels, and a channels-last memory layout, sampling with a first-order Euler integrator. Compliance $\rho_R$ is computed for each sampler configuration against the FDTD reference using Eq.~\eqref{eq:compliance}.

\bibliographystyle{apsrev4-2}
\bibliography{references}

@article{rizzo2023massively,
  title={Massively scalable Kerr comb-driven silicon photonic link},
  author={Rizzo, Anthony and Novick, Asher and Gopal, Vignesh and Kim, Bok Young and Ji, Xingchen and Daudlin, Stuart and Okawachi, Yoshitomo and Cheng, Qixiang and Lipson, Michal and Gaeta, Alexander L and others},
  journal={Nature Photonics},
  volume={17},
  number={9},
  pages={781--790},
  year={2023},
  publisher={Nature Publishing Group UK London},
  doi={https://doi.org/10.1038/s41566-023-01244-7}
}

@article{daudlin2025three,
  title={Three-dimensional photonic integration for ultra-low-energy, high-bandwidth interchip data links},
  author={Daudlin, Stuart and Rizzo, Anthony and Lee, Sunwoo and Khilwani, Devesh and Ou, Christine and Wang, Songli and Novick, Asher and Gopal, Vignesh and Cullen, Michael and Parsons, Robert and others},
  journal={Nature Photonics},
  volume={19},
  number={5},
  pages={502--509},
  year={2025},
  publisher={Nature Publishing Group UK London},
  doi={https://doi.org/10.1038/s41566-025-01633-0}
}

@article{rogers2021universal,
  title={A universal 3D imaging sensor on a silicon photonics platform},
  author={Rogers, Christopher and Piggott, Alexander Y and Thomson, David J and Wiser, Robert F and Opris, Ion E and Fortune, Steven A and Compston, Andrew J and Gondarenko, Alexander and Meng, Fanfan and Chen, Xia and others},
  journal={Nature},
  volume={590},
  number={7845},
  pages={256--261},
  year={2021},
  publisher={Nature Publishing Group UK London},
  doi={https://doi.org/10.1038/s41586-021-03259-y}
}

@article{ahmed2025universal,
  title={Universal photonic artificial intelligence acceleration},
  author={Ahmed, Sufi R and Baghdadi, Reza and Bernadskiy, Mikhail and Bowman, Nate and Braid, Ryan and Carr, Jim and Chen, Chen and Ciccarella, Pietro and Cole, Matthew and Cooke, John and others},
  journal={Nature},
  volume={640},
  number={8058},
  pages={368--374},
  year={2025},
  publisher={Nature Publishing Group UK London},
  doi={https://doi.org/10.1038/s41586-025-08854-x}
}

@article{hua2025integrated,
  title={An integrated large-scale photonic accelerator with ultralow latency},
  author={Hua, Shiyue and Divita, Erwan and Yu, Shanshan and Peng, Bo and Roques-Carmes, Charles and Su, Zhan and Chen, Zhang and Bai, Yanfei and Zou, Jinghui and Zhu, Yunpeng and others},
  journal={Nature},
  volume={640},
  number={8058},
  pages={361--367},
  year={2025},
  publisher={Nature Publishing Group UK London},
  doi={https://doi.org/10.1038/s41586-025-08786-6}
}

@article{vos2007silicon,
  title={Silicon-on-Insulator microring resonator for sensitive and label-free biosensing},
  author={Vos, Katrien De and Bartolozzi, Irene and Schacht, Etienne and Bienstman, Peter and Baets, Roel},
  journal={Optics express},
  volume={15},
  number={12},
  pages={7610--7615},
  year={2007},
  publisher={Optical Society of America},
  doi={https://doi.org/10.1364/OE.15.007610}
}

@article{yu2018silicon,
  title={Silicon-chip-based mid-infrared dual-comb spectroscopy},
  author={Yu, Mengjie and Okawachi, Yoshitomo and Griffith, Austin G and Picqu{\'e}, Nathalie and Lipson, Michal and Gaeta, Alexander L},
  journal={Nature communications},
  volume={9},
  number={1},
  pages={1869},
  year={2018},
  publisher={Nature Publishing Group UK London},
  doi={https://doi.org/10.1038/s41467-018-04350-1}
}

@article{bandyopadhyay2024single,
  title={Single-chip photonic deep neural network with forward-only training},
  author={Bandyopadhyay, Saumil and Sludds, Alexander and Krastanov, Stefan and Hamerly, Ryan and Harris, Nicholas and Bunandar, Darius and Streshinsky, Matthew and Hochberg, Michael and Englund, Dirk},
  journal={Nature Photonics},
  volume={18},
  number={12},
  pages={1335--1343},
  year={2024},
  publisher={Nature Publishing Group UK London},
  doi={https://doi.org/10.1038/s41566-024-01567-z}
}

@article{sun2015single,
  title={Single-chip microprocessor that communicates directly using light},
  author={Sun, Chen and Wade, Mark T and Lee, Yunsup and Orcutt, Jason S and Alloatti, Luca and Georgas, Michael S and Waterman, Andrew S and Shainline, Jeffrey M and Avizienis, Rimas R and Lin, Sen and others},
  journal={Nature},
  volume={528},
  number={7583},
  pages={534--538},
  year={2015},
  publisher={Nature Publishing Group UK London},
  doi={https://doi.org/10.1038/nature16454}
}

@book{TafloveHagness2005FDTD,
  author    = {Allen Taflove and Susan C. Hagness},
  title     = {Computational Electrodynamics: The Finite-Difference Time-Domain Method},
  edition   = {3rd},
  publisher = {Artech House},
  year      = {2005}
}

@article{Teixeira2023FDTDMethods,
  author  = {F. L. Teixeira and C. Sarris and Y. Zhang and D.-Y. Na and J.-P. Berenger and Y. Su and M. Okoniewski and W. C. Chew and V. Backman and J. J. Simpson},
  title   = {Finite-Difference Time-Domain Methods},
  journal = {Nature Reviews Methods Primers},
  volume  = {3},
  pages   = {75},
  year    = {2023},
  doi     = {10.1038/s43586-023-00257-4}
}

@article{Oskooi2010Meep,
  author  = {Ardavan F. Oskooi and David Roundy and Mihai Ibanescu and Peter Bermel and J. D. Joannopoulos and Steven G. Johnson},
  title   = {Meep: A flexible free-software package for electromagnetic simulations by the FDTD method},
  journal = {Computer Physics Communications},
  volume  = {181},
  number  = {3},
  pages   = {687--702},
  year    = {2010},
  doi     = {10.1016/j.cpc.2009.11.008}
}

@inproceedings{BurrFarjadpour2005FDTDAccuracy,
  author    = {Geoffrey W. Burr and Ardavan Farjadpour},
  title     = {Balancing accuracy against computation time: 3-{D} {FDTD} for nanophotonics device optimization},
  booktitle = {Proceedings of SPIE},
  volume    = {5733},
  pages     = {336--347},
  year      = {2005},
  doi       = {https://doi.org/10.1117/12.590732}
}

@book{ReedKnights2004SiliconPhotonics,
  author    = {Graham T. Reed and Andrew P. Knights},
  title     = {Silicon Photonics: An Introduction},
  publisher = {John Wiley \& Sons},
  year      = {2004},
  doi       = {10.1002/0470014180}
}

@misc{Lipman2023FlowMatching,
  title={Flow matching for generative modeling},
  author={Lipman, Yaron and Chen, Ricky TQ and Ben-Hamu, Heli and Nickel, Maximilian and Le, Matt},
  journal={arXiv preprint arXiv:2210.02747},
  year={2022},
  eprint={2210.02747},
  archivePrefix = {arXiv},
  primaryClass  = {cs.LG},
  doi={https://doi.org/10.48550/arXiv.2210.02747}
}

@inproceedings{Liu2023RectifiedFlow,
  author    = {Xingchao Liu and Chengyue Gong and Qiang Liu},
  title     = {Flow Straight and Fast: Learning to Generate and Transfer Data with Rectified Flow},
  booktitle = {International Conference on Learning Representations (ICLR)},
  year      = {2023}
}

@misc{Baldan2025PBFM,
  author        = {Giacomo Baldan and Qiang Liu and Alberto Guardone and Nils Thuerey},
  title         = {Physics vs Distributions: Pareto Optimal Flow Matching with Physics Constraints},
  year          = {2025},
  eprint        = {2506.08604},
  archivePrefix = {arXiv},
  primaryClass  = {cs.LG},
  doi           = {10.48550/arXiv.2506.08604}
}

@article{Raissi2019PINNs,
  author  = {Maziar Raissi and Paris Perdikaris and George Em Karniadakis},
  title   = {Physics-informed neural networks: A deep learning framework for solving forward and inverse problems involving nonlinear partial differential equations},
  journal = {Journal of Computational Physics},
  volume  = {378},
  pages   = {686--707},
  year    = {2019},
  doi     = {10.1016/j.jcp.2018.10.045}
}

@misc{Li2021FNO,
  title={Fourier neural operator for parametric partial differential equations},
  author={Li, Zongyi and Kovachki, Nikola and Azizzadenesheli, Kamyar and Liu, Burigede and Bhattacharya, Kaushik and Stuart, Andrew and Anandkumar, Anima},
  eprint        = {2010.08895},
  archivePrefix = {arXiv},
  primaryClass  = {cs.LG},
  year={2020},
  doi={https://doi.org/10.48550/arXiv.2010.08895}
}

@article{Lu2021DeepONet,
  author  = {Lu Lu and Pengzhan Jin and Guofei Pang and Zhongqiang Zhang and George Em Karniadakis},
  title   = {Learning nonlinear operators via DeepONet based on the universal approximation theorem of operators},
  journal = {Nature Machine Intelligence},
  volume  = {3},
  pages   = {218--229},
  year    = {2021},
  doi     = {10.1038/s42256-021-00302-5}
}

@article{Jiang2021DeepNNPhotonicDevices,
  author  = {Jiaqi Jiang and Mingkun Chen and Jonathan A. Fan},
  title   = {Deep neural networks for the evaluation and design of photonic devices},
  journal = {Nature Reviews Materials},
  volume  = {6},
  pages   = {679--700},
  year    = {2021},
  doi     = {10.1038/s41578-020-00260-1}
}

@article{Tahersima2019DNNSplitters,
  author  = {Mohammad H. Tahersima and Keisuke Kojima and Toshiaki Koike-Akino and Devesh Jha and Bingnan Wang and Chungwei Lin and Kieran Parsons},
  title   = {Deep Neural Network Inverse Design of Integrated Photonic Power Splitters},
  journal = {Scientific Reports},
  volume  = {9},
  pages   = {1368},
  year    = {2019},
  doi     = {10.1038/s41598-018-37952-2}
}

@article{Tang2020GenerativeLPR,
  author  = {Yingheng Tang and Keisuke Kojima and Toshiaki Koike-Akino and Ye Wang and Pengxiang Wu and Youye Xie and Mohammad H. Tahersima and Devesh K. Jha and Kieran Parsons and Minghao Qi},
  title   = {Generative Deep Learning Model for Inverse Design of Integrated Nanophotonic Devices},
  journal = {Laser \& Photonics Reviews},
  volume  = {14},
  number  = {12},
  pages   = {2000287},
  year    = {2020},
  doi     = {10.1002/lpor.202000287}
}

@article{Piggott2015DemultiplexerNatPhoton,
  author  = {Alexander Y. Piggott and Jesse Lu and Konstantinos G. Lagoudakis and Jan Petykiewicz and Thomas M. Babinec and Jelena Vu{\v{c}}kovi{\'c}},
  title   = {Inverse design and demonstration of a compact and broadband on-chip wavelength demultiplexer},
  journal = {Nature Photonics},
  volume  = {9},
  pages   = {374--377},
  year    = {2015},
  doi     = {10.1038/nphoton.2015.69}
}

@article{Piggott2017FabricationConstrainedSciRep,
  author  = {Alexander Y. Piggott and Jan Petykiewicz and Logan Su and Jelena Vu{\v{c}}kovi{\'c}},
  title   = {Fabrication-constrained nanophotonic inverse design},
  journal = {Scientific Reports},
  volume  = {7},
  pages   = {1786},
  year    = {2017},
  doi     = {10.1038/s41598-017-01939-2}
}

@article{Yeung2023DeepAdjoint,
  author  = {Christopher Yeung and Benjamin Pham and Ryan Tsai and Katherine T. Fountaine and Aaswath P. Raman},
  title   = {DeepAdjoint: An All-in-One Photonic Inverse Design Framework Integrating Data-Driven Machine Learning with Optimization Algorithms},
  journal = {ACS Photonics},
  volume  = {10},
  number  = {4},
  pages   = {884--891},
  year    = {2023},
  doi     = {10.1021/acsphotonics.2c00968}
}

@article{Pan2023PhotonicsReview,
  author  = {Zongyong Pan and Xiaomin Pan},
  title   = {Deep Learning and Adjoint Method Accelerated Inverse Design in Photonics: A Review},
  journal = {Photonics},
  volume  = {10},
  number  = {7},
  pages   = {852},
  year    = {2023},
  doi     = {10.3390/photonics10070852}
}

@article{Molesky2018InverseDesign,
  author  = {Sean Molesky and Zin Lin and Alexander Y. Piggott and Weiliang Jin and Jelena Vu{\v{c}}kovi{\'c} and Alejandro W. Rodriguez},
  title   = {Inverse design in nanophotonics},
  journal = {Nature Photonics},
  volume  = {12},
  pages   = {659--670},
  year    = {2018},
  doi     = {10.1038/s41566-018-0246-9}
}

@inproceedings{Ronneberger2015UNet,
  author    = {Olaf Ronneberger and Philipp Fischer and Thomas Brox},
  title     = {U-Net: Convolutional Networks for Biomedical Image Segmentation},
  booktitle = {Medical Image Computing and Computer-Assisted Intervention (MICCAI)},
  series    = {Lecture Notes in Computer Science},
  volume    = {9351},
  pages     = {234--241},
  publisher = {Springer},
  year      = {2015},
  doi       = {10.1007/978-3-319-24574-4_28}
}

@article{Chen2022WaveYNet,
  author  = {Mingkun Chen and Robert Lupoiu and Chenkai Mao and Der-Han Huang and Jiaqi Jiang and Philippe Lalanne and Jonathan A. Fan},
  title   = {High Speed Simulation and Freeform Optimization of Nanophotonic Devices with Physics-Augmented Deep Learning},
  journal = {ACS Photonics},
  volume  = {9},
  number  = {9},
  pages   = {3110--3123},
  year    = {2022},
  doi     = {10.1021/acsphotonics.2c00876}
}

@article{Lim2022MaxwellNet,
  author  = {Joowon Lim and Demetri Psaltis},
  title   = {MaxwellNet: Physics-driven deep neural network training based on Maxwell's equations},
  journal = {APL Photonics},
  volume  = {7},
  number  = {1},
  pages   = {011301},
  year    = {2022},
  doi     = {10.1063/5.0071616}
}

@article{Ma2021DLPhotonicsReview,
  author  = {Wei Ma and Zhaocheng Liu and Zhaxylyk A. Kudyshev and Alexandra Boltasseva and Wenshan Cai and Yongmin Liu},
  title   = {Deep learning for the design of photonic structures},
  journal = {Nature Photonics},
  volume  = {15},
  pages   = {77--90},
  year    = {2021},
  doi     = {10.1038/s41566-020-0685-y}
}

@inproceedings{Song2021ScoreSDE,
  author    = {Yang Song and Jascha Sohl-Dickstein and Durk P. Kingma and Abhishek Kumar and Stefano Ermon and Ben Poole},
  title     = {Score-Based Generative Modeling through Stochastic Differential Equations},
  booktitle = {International Conference on Learning Representations (ICLR)},
  year      = {2021}
}

@inproceedings{Huang2024DiffusionPDE,
  author    = {Jiahe Huang and Guandao Yang and Zichen Wang and Jeong Joon Park},
  title     = {DiffusionPDE: Generative PDE-Solving Under Partial Observation},
  booktitle = {Advances in Neural Information Processing Systems (NeurIPS)},
  volume    = {37},
  pages     = {130291--130323},
  year      = {2024}
}

\end{document}